%% file: paper-arxiv-cityseer-package.tex
\documentclass{article}

\input{shared/arxivSettings}

\title{The cityseer Python package for pedestrian-scale network-based urban analysis}

\date{} 

\input{shared/author.tex}

\renewcommand{\shorttitle}{The cityseer Python package}

\hypersetup{
	pdftitle={
		The cityseer Python package for pedestrian-scale network-based urban analysis
	},
	pdfsubject={
		physics.soc-ph
	},
	pdfauthor={
		Gareth~D.~Simons,
	},
	pdfkeywords={
		computation,
		data science,
		GIS,
		land-use analysis,
		morphometrics,
		network analysis,
		spatial analysis,
		urban analytics,
		urban planning,
		urban morphology,
		urbanism
	},
}

\begin{document}
\maketitle
\begin{abstract}
	\input{content/0_abstract.tex}
\end{abstract}
\keywords{
	computation
	\and data-science
	\and GIS
	\and land-use analysis
	\and morphometrics
	\and network analysis
	\and spatial analysis
	\and urban analytics
	\and urban planning
	\and urban morphology
	\and urbanism
}
\input{content/1_overview.tex}
\input{content/2_localised_urban_analysis.tex}
\input{content/3_localised_methods.tex}
\input{content/4_design_decisions.tex}
\input{content/5_prototypical_workflow.tex}
\input{content/6_summary.tex}

\section{Acknowledgements}
\input{shared/acknowledge_phd.tex}

\section{Citations}
\printbibliography[heading=none]{}
\end{document}

%% file: shared/arxivSettings.tex
\usepackage{./shared/arxiv}
\usepackage[natbib=true,style=authoryear,doi=true,isbn=true,url=false,eprint=true]{biblatex}

\usepackage[utf8]{inputenc} 
\usepackage[T1]{fontenc}    
\usepackage{hyperref}       
\usepackage{url}            
\usepackage{booktabs}       
\usepackage{amsfonts}       
\usepackage{nicefrac}       
\usepackage{microtype}      
\usepackage{graphicx}

\usepackage{doi}

\usepackage{gensymb} 
\usepackage{rotating} 
\usepackage{caption}
\usepackage{subcaption}
\usepackage[nobottomtitles*]{titlesec}
\usepackage{amsmath}
\usepackage{setspace}

\newcommand{\code}[1]{\lstinline[language=bash, basicstyle=\ttfamily\small]|#1|}
\usepackage{color}
\definecolor{lightgrey}{rgb}{0.975, 0.975, 0.975}
\definecolor{midgrey}{rgb}{0.6, 0.6, 0.6}
\definecolor{deepgrey}{rgb}{0.2, 0.2, 0.2}
\definecolor{codegreen}{rgb}{0, 0.7, 0}
\definecolor{codepink}{rgb}{0.8196, 0.10, 0.654}
\usepackage{listings}

%% file: shared/author.tex
\author{
	\href{https://orcid.org/0000-0003-3790-0638}{
		\includegraphics[width=0.25cm, height=0.25cm, keepaspectratio]{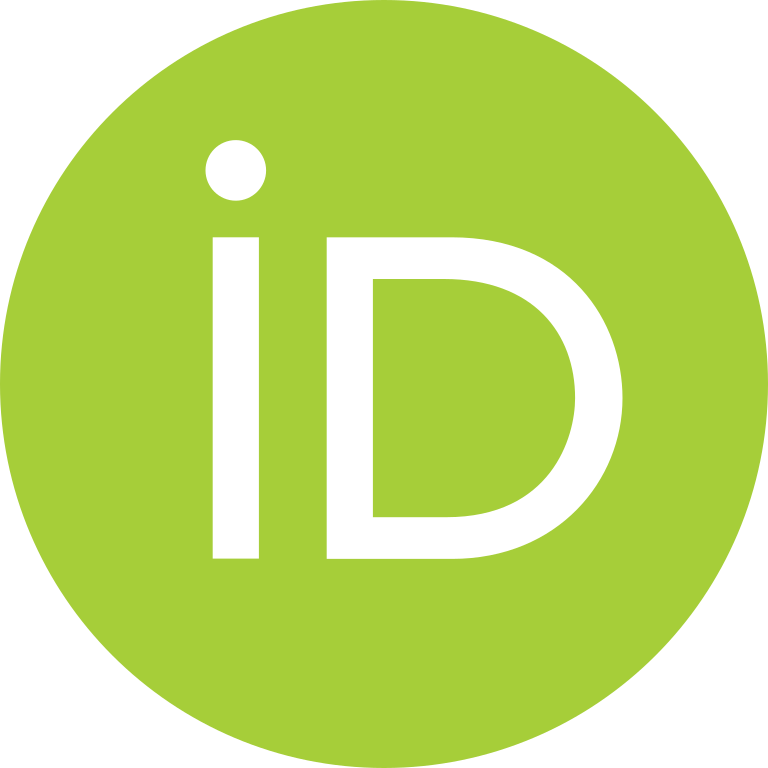}
		\hspace{1mm}Gareth D. Simons
	}
	\thanks{Benchmark Urbanism \texttt{gareth@benchmarkurbanism.com}}
}

%% file: content/0_abstract.tex
\code{cityseer-api} is a \code{Python} package consisting of computational tools for fine-grained street-network and land-use analysis, helpful in assessing the morphological precursors to vibrant neighbourhoods. It is underpinned by network-based methods developed specifically for urban analysis at the pedestrian scale. \code{cityseer-api} computes a variety of node and segment-based network centrality methods, land-use accessibility and mixed-use measures, and statistical aggregations. Accessibilities and aggregations are computed dynamically over the street-network while taking walking distance thresholds and the direction of approach into account, and can optionally incorporate spatial impedances and network decomposition to increase spatial precision. The use of \code{Python} facilitates compatibility with popular computational tools for network manipulation (\code{NetworkX}), geospatial topology (\code{shapely}), geospatial data state management (\code{GeoPandas}), and the \code{NumPy} stack of scientific packages. The provision of robust network cleaning tools aids the use of OpenStreetMap data for network analysis. Underlying loop-intensive algorithms are implemented in \code{Numba} \code{JIT} compiled code so that the methods scale efficiently to larger cities and regions.

Online documentation is available from \href{https://cityseer.benchmarkurbanism.com}{cityseer.benchmarkurbanism.com}, and the Github repository is available at \href{https://github.com/benchmark-urbanism/cityseer}{github.com/benchmark-urbanism/cityseer}. Example notebooks are available at \href{https://cityseer.benchmarkurbanism.com/examples/}{cityseer.benchmarkurbanism.com/examples/}.

%% file: content/1_overview.tex
\section{Overview}

\code{cityseer-api} is a street-network-based spatial analysis package for pedestrian-scale urban morphological analysis. It combines street-network decomposition, distance-weighted implementations of network centrality and mixed-use measures, and contextually sensitive distance and aggregational methods to generate observations with a high degree of spatial precision. The synthesis of these techniques enables \code{cityseer-api} to describe variances in morphological metrics at regular intervals along streetfronts in a manner facilitating comparative analysis of walkability, mixed-uses, and land-use accessibilities for urban planning scenarios.

Centrality and land-use analysis methods rely extensively on shortest-path algorithms, presenting substantial computational complexity due to nested computational loops. Pure \code{Python} network-based measures, such as those implemented in \code{NetworkX} \citep{Hagberg2008} are consequently prohibitively slow if applied to analysis for larger towns and cities. Performance improvements can be attained through use of packages such as \code{Graph-Tool} \citep{Peixoto2014}, \code{igraph} \citep{Csardi2006}, \code{depthmapX} \citep{depthmapx_development_team_depthmapx_2017}, or \code{pandana} \citep{foti_generalized_2012}, which wrap underlying optimised \code{C} or \code{C++} code. However, reliance on packages underpinned by lower-level programming languages presents a challenge for explorative research because it becomes difficult to manipulate underlying algorithms without incurring complexity or a loss of computational efficiency. This conundrum prompted the development of the codebase formalised as \code{cityseer-api}, which has adopted an approach leveraging pure \code{Python} and \code{NumPy} \citep{Harris2020}, but with computationally intensive loops optimised through use of \code{Numba} \code{JIT} (just-in-time) compilation \citep{Lam2015}. This approach has allowed for wide-ranging experimentation while permitting a set of pertinent issues to be addressed:
\begin{itemize}
      \item \code{cityseer-api} employs a `moving-window' form of spatial analysis: each node in the network is visited in turn, with the network then isolated at a range of specified walking distance thresholds from the currently selected node. Centrality, land-use, and aggregational methods can then be computed for the locally windowed context. This is similar to radial forms of analysis used in Space Syntax (in the context of street-network centralities) and the notion of `overlapping buffer queries' used in \code{pandana} (within the context of land-use accessibilities and data aggregations). Such forms of windowed distance thresholds can be based on either crow-flies euclidean distances or true network distances \citep{Cooper2015}; \code{cityseer-api} takes the position that true network distances are most representative when working at smaller pedestrian distance thresholds, particularly when applied to land-use accessibilities and mixed-use calculations. Moving window analysis is advantageous because it clearly and consistently defines the network boundary in relation to the current point of analysis, therefor sidestepping issues such as the robust definition of town or city boundaries, prevention of edge rolloff effects, and difficulties regarding normalisation of measures for comparisons between locations on differently sized networks.
      \item It is common to use either shortest-distance or simplest-path (least angular `distance') impedance heuristics when computing network centralities. When using simplest-path heuristics, it is necessary to modify the underlying shortest-path algorithms to prevent side-stepping of sharp angular turns; otherwise, two smaller side-steps can be combined to `short-cut' sharp corners \citep{Turner2007}. This safeguard is not available in off-the-shelf network analysis packages. It is common for centrality methods to be applied to either primal network representations, generally used with shortest-path methods such as those applied by \emph{multiple centrality assessment} analysis \citep{Porta2006}, or dual network representations, typically used with simplest-path methods in the tradition of \emph{space syntax} \citep{Hillier1984}. \code{cityseer-api} incorporates both forms of analysis while also allowing for angular centralities to be calculated on primal networks so that topological divergences between primal and dual networks do not skew observations comparing shortest and simplest path heuristics.
      \item A range of centrality and mixed-use methods is available for urban analysis; \code{cityseer-api} incorporates specialised forms of these methods including distance-weighted versions greatly accentuating spatial precision. Some conventional methods, even if widely used, can be problematic for urban analysis workflows: specifically, conventional formulations of closeness centrality do not behave as anticipated for windowed networks, and mixed-use analysis methods derived from larger-scale zoned or gridded aggregations become problematic if interpreted within the context of streets. \code{cityseer-api} incorporates implementations of these methods that are not susceptible to these issues. These methods and their implications are developed and explored at length in the accompanying papers on network centrality methods \citep{Simons2021c} and mixed-use methods \citep{Simons2021d};
      \item Centrality methods are susceptible to topological distortions arising from `messy' network representations as well as due to the conflation of topological and geometrical properties of street-networks, which has a detrimental impact on the calculation of network centralities. \code{cityseer-api} addresses these through the inclusion of robust network cleaning functions with substantial effort directed towards procedures for splitting geometrical properties from topological representations; the removal of parallel roadways; and the inclusion of segmentised forms of centrality measures, which are less susceptible to distortions introduced by varying intensities of nodes;
      \item Pedestrian-scale analysis requires approaches facilitating the evaluation of respective measures at finely-spaced intervals along street fronts. Further, granular evaluation of land-use accessibilities and mixed-uses requires that land-uses be assigned to the street-network in a contextually precise manner. These are addressed in \code{cityseer-api} through the application of a network decomposition technique. Instead of assigning data-points to the nearest node, \code{cityseer-api} searches for the closest adjacent street edge and then uses a bidirectional assignment method. This allows for distances in relation to aggregations or accessibilities to be computed dynamically, while taking into account the direction of approach from the currently windowed node.
\end{itemize}

\begin{figure}[htbp]
      \centering
      \includegraphics[width=\textwidth, height=10cm, keepaspectratio]{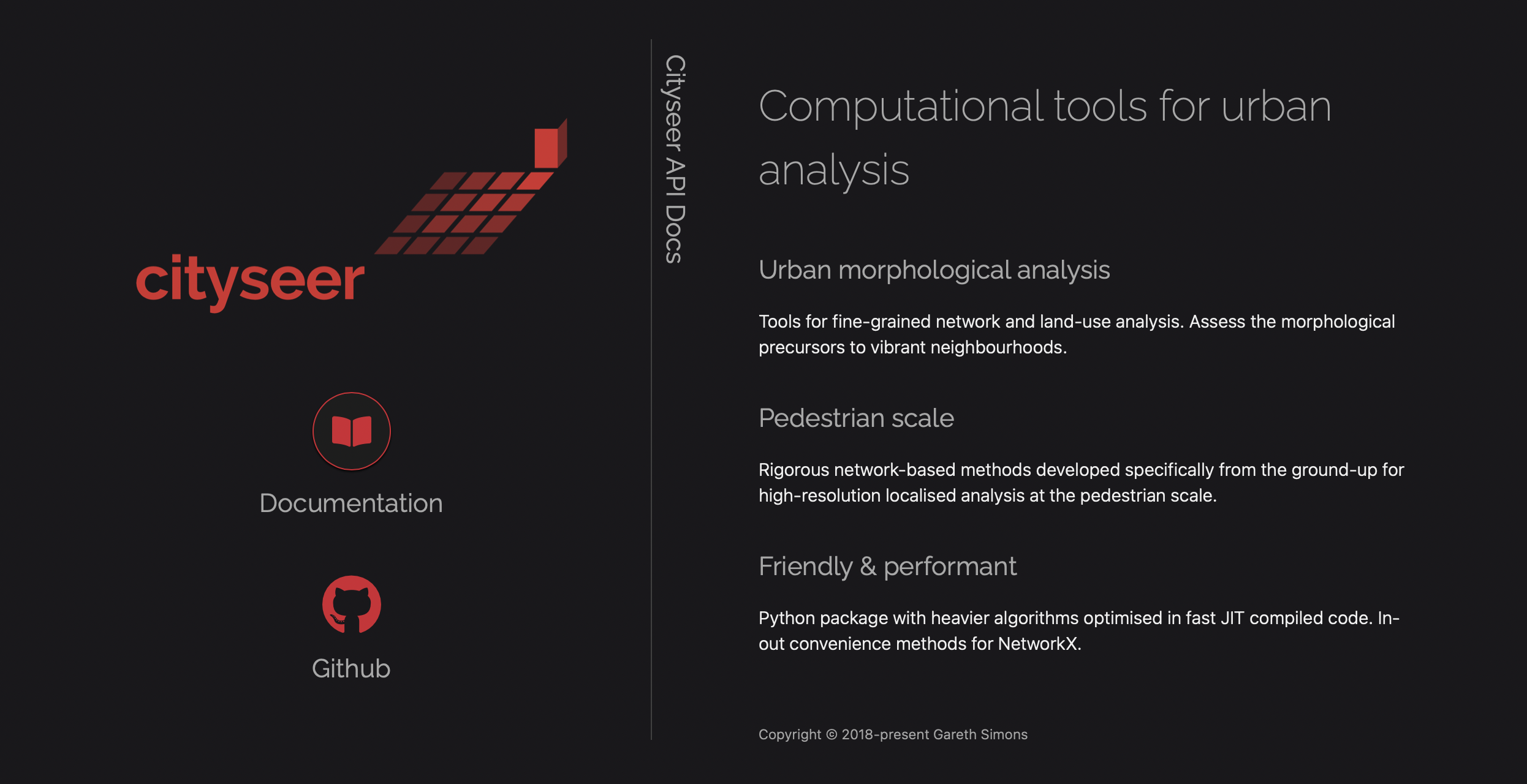}
      \caption{\code{cityseer-api} documentation homepage.}\label{fig:cityseer_docs_1}
\end{figure}

\begin{figure}[htbp]
      \centering
      \includegraphics[width=\textwidth, height=10cm, keepaspectratio]{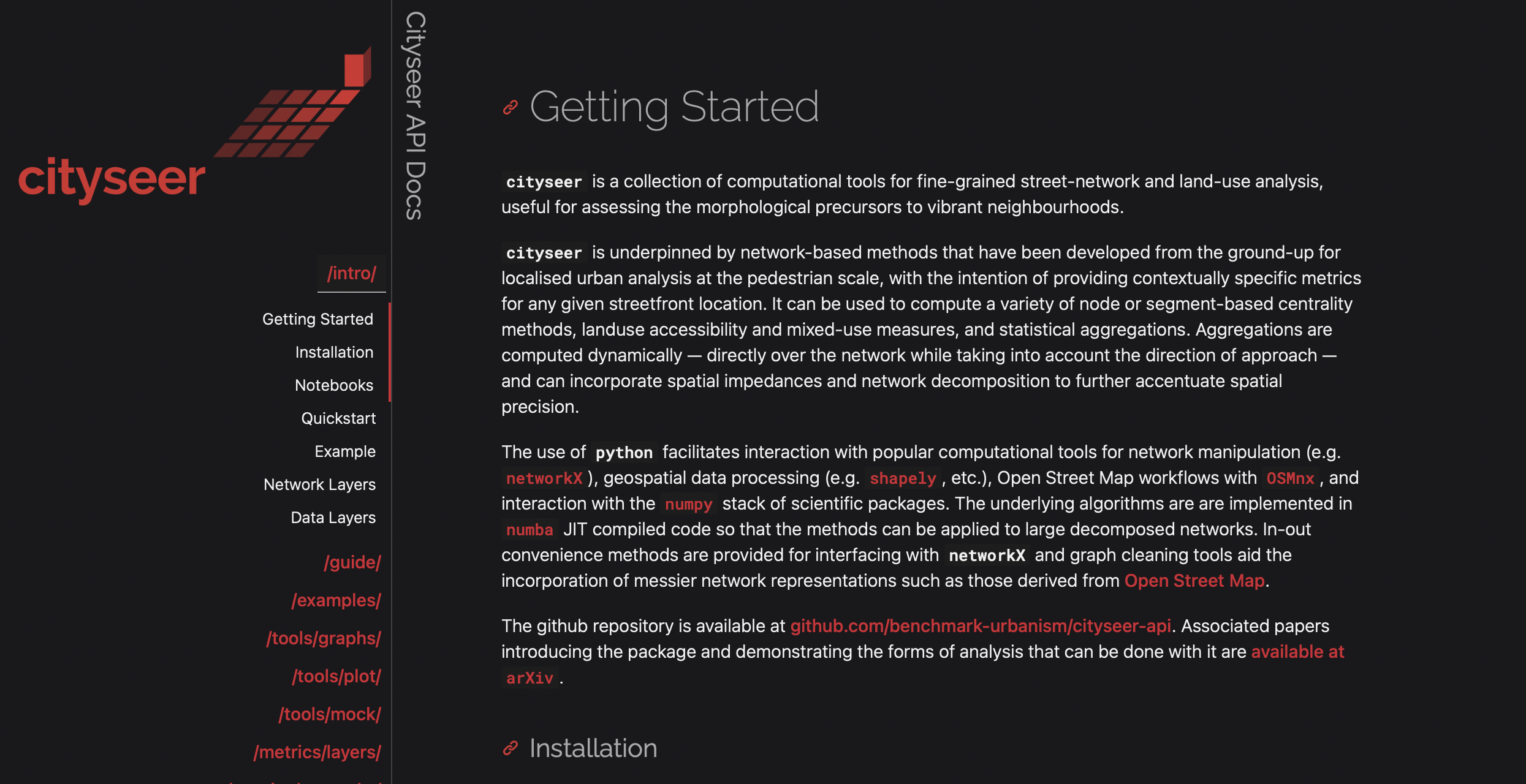}
      \caption{Getting started guide from the \code{cityseer-api} documentation website.}\label{fig:cityseer_docs_2}
\end{figure}

\code{cityseer-api} is intended to be data source agnostic and includes convenience methods for the general preparation of networks and their conversion into (and out of) the lower-level data structures used by the underlying algorithms. These network utility methods are designed to work with \code{NetworkX} to facilitate ease of use, and to enable workflows incorporating data from sources such as \code{PostGIS} or OpenStreetMap data, whether from API queries or from \code{OSMnx} \citep{Boeing2017}. Data state is managed with \code{GeoPandas} \citep{jordahl_geopandasgeopandas_2020} to facilitate downstream analysis and modelling and for bridging to \code{GeoPandas} based workflows, such as used by \code{momepy} \citep{Fleischmann2019}.

Detailed package documentation is available at \href{https://cityseer.benchmarkurbanism.com}{cityseer.benchmarkurbanism.com} (Figures~\ref{fig:cityseer_docs_1} and~\ref{fig:cityseer_docs_2}), including a \href{https://cityseer.benchmarkurbanism.com/guide/}{guide} and a growing collection of \href{https://cityseer.benchmarkurbanism.com/examples/}{examples}. Discussion and examples relating to use with other packages is provided in the documentation \href{https://cityseer.benchmarkurbanism.com/guide/}{guide}

A complement of code formatters, linters, type-checkers, and unit tests maintains the integrity of the code-base through general package maintenance and upgrade cycles. Where feasible, centrality methods are checked against \code{NetworkX} or against manually checked testing scenarios. Extensive mock data and test plots have been used to visually confirm the intended behaviour for divergent simplest and shortest-path heuristics and for confirming the assignment and aggregation of data-points.

%% file: content/2_localised_urban_analysis.tex
\section{Nuances of spatial aggregation}\label{localised-computational-methods}

Computational tools have dramatically increased the range and depth of scientific analysis. Likewise, methods applied to spatial analysis have been revolutionised and hold tremendous potential for rigorous and scalable forms of urban analysis, which may prove helpful as benchmarking tools for principles espoused in urban theory and policy. Nevertheless, computational constraints and coarse data sources have historically favoured the use of larger areal units of spatial aggregation, such as spatial grids or areal zones, and these have typically been combined with simplified proximity methods such as crow-flies instead of network-based distance measures \citep{Logan2017, Araldi2016}.

\begin{figure}[htbp]
  \centering
  \includegraphics[width=\textwidth, keepaspectratio]{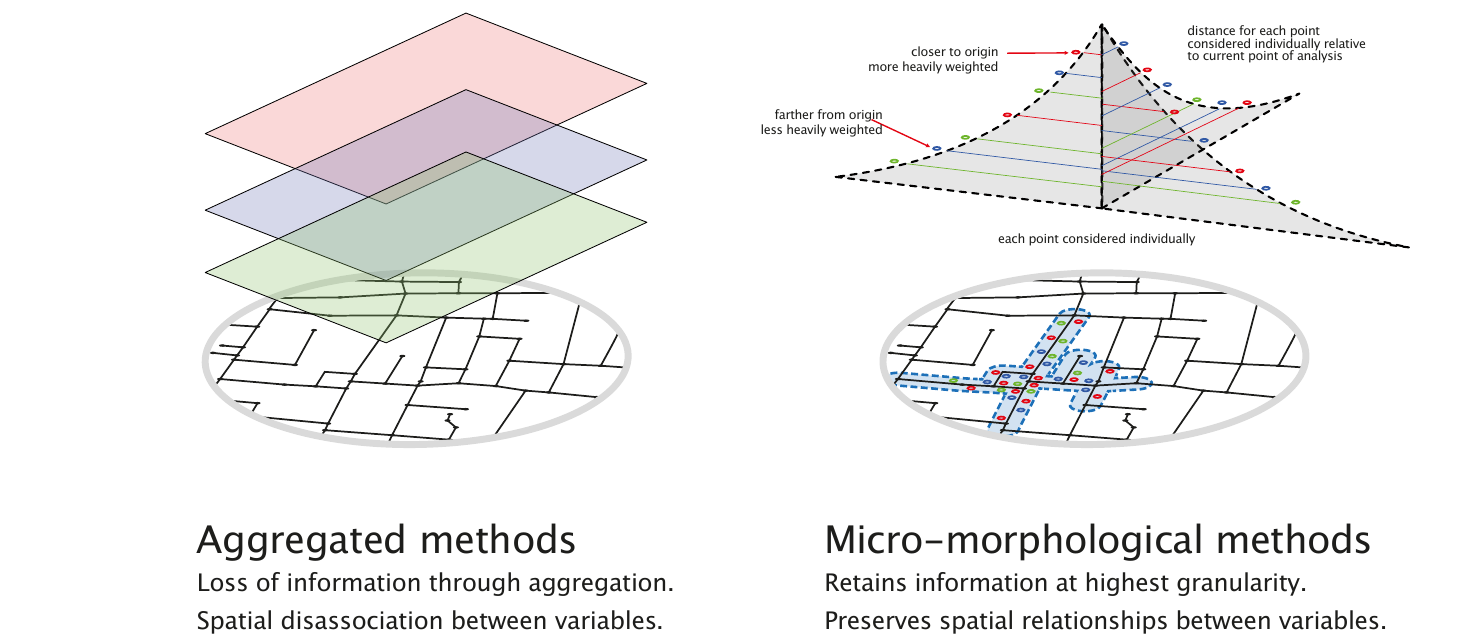}
  \caption[Overarching spatial units as opposed to a moving window (localised) approach.]{Overarching spatial analysis methods, where variables are aggregated to encompassing grids or zones (left), do not work well for purposes of urban analysis from the perspective of architecture and urbanism. This is because overarching methods collapse spatial information into larger-scale aggregations with the implication that data is no longer contextually relevant to a pedestrian's perception of space as defined by streetfronts. Unlike overarching aggregations, locally windowed methods (right) do not collapse data-points and do not discard relationships between variables relative to a selected location on the street-network. Locations and distances remain anchored relative to street-networks and walking tolerances from a given point in space.}\label{fig:collapse}
\end{figure}

The use of crude spatial aggregations presents a dilemma from the perspective of urban design. The averaging of variables to overarching spatial units causes the loss of higher-resolution information along streetfronts while obscuring relationships between observed variables for individual data-points, buildings, or plots. Statistical forms of analysis consequently encounter the \emph{Ecological Fallacy}: correlations which may have been valid for a larger unit of analysis can become misleading if interpreted within the disaggregated context \citep{Robinson2009}; expressed differently, the context of a local street corner may be significantly removed from larger-scale statistical averages for a given neighbourhood, city, or region. More generally, aggregations can mask confounding variables (\emph{Simpson's Paradox}) and the loss of information across `geography' (space) and `history' (time) may similarly confound spatially aggregated data \citep{Cressie1998}. This segues into the \emph{Modifiable Areal Unit Problem} (MAUP) where statistical observations derived from spatially aggregated data is sensitive to the scale of aggregation; the arrangement of the data in relation to zonal extents; and spatial autocorrelation in the variables. As a rule-of-thumb, larger aggregations increase sensitivity to MAUP because of a smoothing effect in the distribution of the data due to decreasing levels of variance, with the implication that correlation coefficients will strengthen as the unit of aggregation increases \citep{Fotheringham1991}. Variance is likewise affected by spatial autocorrelation of variables or by the movement of boundaries relative to the geographic locations of data-points. Different spatial aggregations therefor trigger fluctuating and, sometimes, questionable statistical inferences if applied or interpreted across different scales of analysis or between varied zonal configurations \citep{ROBINSON1956, Thomas1965}. These forms of problem are inherent to the use of spatially aggregated data, and no simple solutions exist, with the issue proving particularly intractable for multivariate analysis. Nevertheless, attempts persist at better defining and managing the issue \citep{Reynolds1998, Duque2018}.

The expanding availability of spatially granular data sources combined with growing access to computational resources has begun to tip the scales in favour of higher-resolution workflows capable of more contextually precise forms of spatial analysis \citep{Yamada2010} that are less susceptible to aggregational artefacts. Further, rich data sources synthesised with street-network-based strategies heralds a paradigm shift from the aerial vantage point of the plan --- traditionally the frame of reference for morphological analysis --- to that of localised pedestrian-centric methods applied directly over the street-network \citep{Araldi2019}: the pedestrian's vantage point can, in a literal sense, become the anchor and point of departure for spatial analysis.

%% file: content/3_localised_methods.tex
\section{Localised methods}\label{localised-methods}

\begin{figure}[htbp]
  \centering
  \includegraphics[width=\textwidth, keepaspectratio]{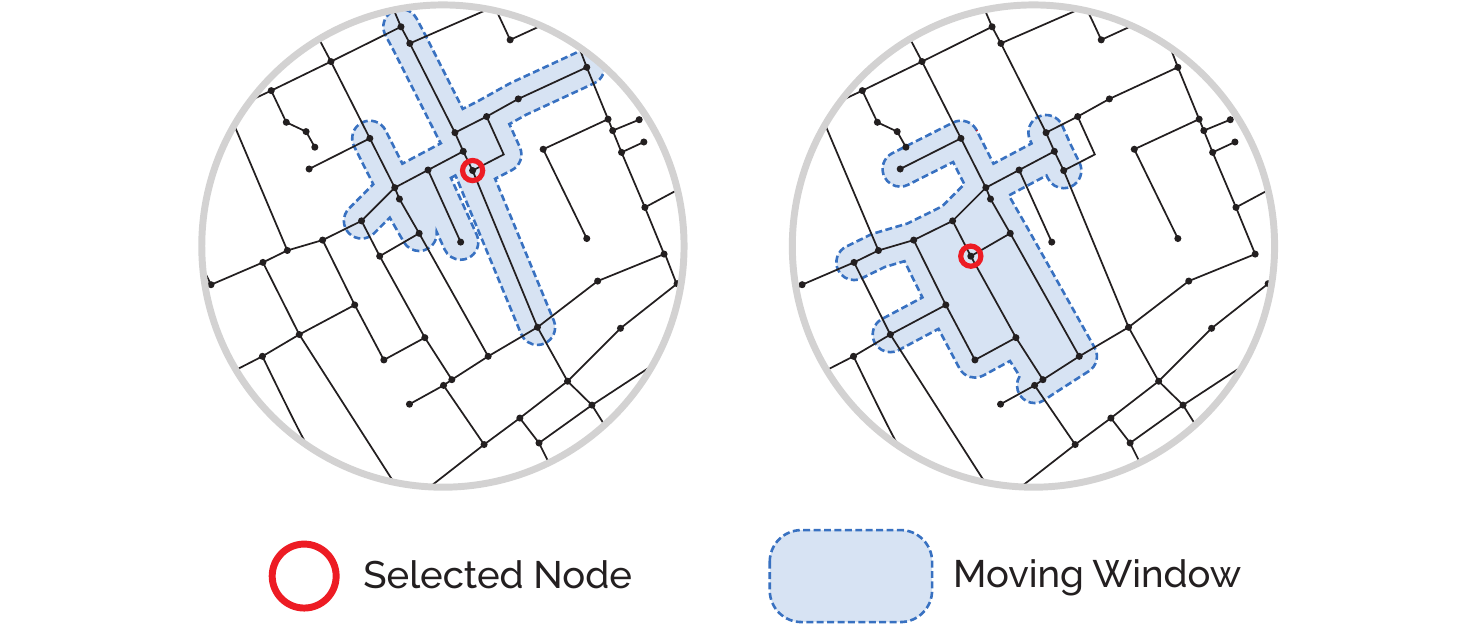}
  \caption[`Moving-window' localised network analysis.]{Moving-window' localised network analysis: metrics are calculated for each node in the network using a `moving-window', taking into account all other nodes or data-points within a specified distance threshold.}\label{fig:moving_window}
\end{figure}

`Moving-window' forms of a localised spatial analysis (also called `radial' or `buffered' methods) differ from gridded or zonal aggregations frequently used in Geographical Information Systems more widely. Calculations and aggregations are unfurled directly over the street-network at a set of selected distance thresholds: an algorithm visits each node in the network in turn; isolates the surrounding nodes at the specified distance thresholds; then centrality, land-use, or other aggregational or statistical measures can be computed for the currently selected location. The process subsequently repeats for every other node in the network, making it a `localised' method because the calculations are repeated on an individual basis relative to each node (Figure~\ref{fig:moving_window}). The full resolution of the data thus remains available to each sampled point: data remains spatially anchored and the distance from each point of analysis to each surrounding data-point is knowable (Figure~\ref{fig:collapse}). Contextually specific relationships between variables are therefor not sacrificed and it becomes possible to use spatial impedances to further accentuate locality relative to pedestrian walking tolerances. The dynamic nature of localised moving-window methods ameliorates the zonal aspect of the Modifiable Areal Unit Problem (MAUP) because the extents are defined consistently and methodically relative to the origin of each point of analysis and the scale of aggregation. Nevertheless, as with MAUP effects more generally, statistical variances tend to decrease for aggregations at increasingly large network distances, with the implication that correlations cannot be directly compared between smaller and larger distance thresholds.

\begin{figure}[htbp]
  \centering
  \includegraphics[width=\textwidth, keepaspectratio]{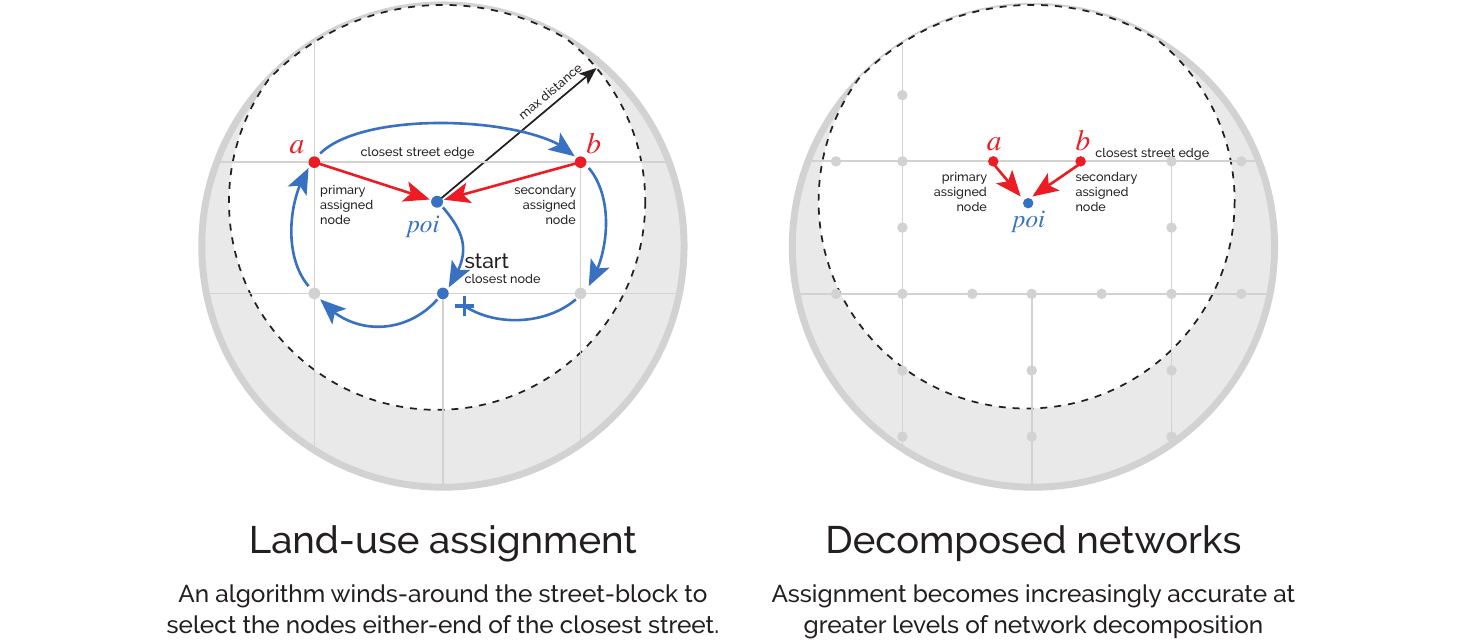}
  \caption[Assignment of land-use datapoints to the street-network.]{Datapoints representing land-uses are assigned to the two network nodes on either side of the nearest adjacent street edge. Therefor, accurate distances can be determined from each point of analysis to each datapoint via the closest adjacent node on the street-network while taking the direction of the approach into account. The algorithm `winds' around the street-network to encircle the datapoint of interest to identify the closest adjacent street edge. The assignment increases in accuracy with increasing levels of network decomposition. In \code{cityseer-api}, the distance from the windowed node to a selected data-point includes the distance from the primary / secondary nodes to the datapoint's actual location in space.}\label{fig:poi_assignment}
\end{figure}

\begin{figure}[htbp]
  \centering
  \includegraphics[width=\textwidth, keepaspectratio]{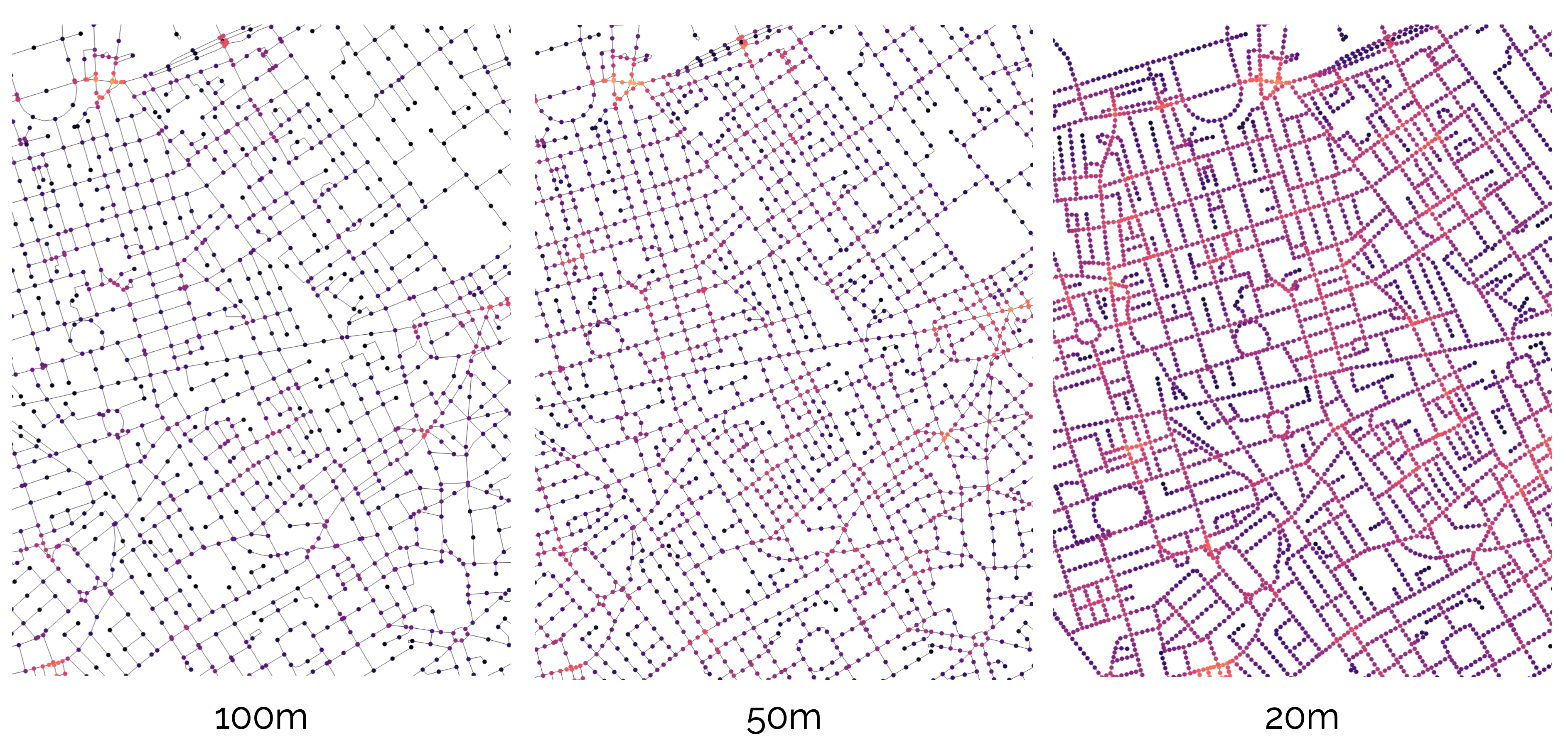}
  \caption[Varying levels of street network decomposition.]{Varying levels of street network decomposition: Increasing the level of decomposition permits a higher resolution of analysis showing the variability of measures along street-fronts, allows for more precise assignments of land-uses to adjacent street segments, and reduces the potential loss of information where longer street segments might otherwise be discarded if intersecting a distance threshold.}\label{fig:decomposition}
\end{figure}

Whereas aggregations at increasingly small distance thresholds become more locally focussed, there is a point at which distance cutoffs can become too small in relation to the topological structure of the network, potentially causing calculations or aggregations to become haphazard because longer street segments increasingly intersect or `overshoot' smaller distance thresholds. These scenarios can cause a spatial `vacuum' of information at the periphery of the moving-window and may consequently exacerbate variances in spatial aggregations, with ensuing artefacts in the distributions and a weakening of correlations. Another problem with longer street segments is that observations are not always sufficiently granular to provide intervening information between widely-spaced nodes on either side of longer street segments. \citet{Yamada2010} applied a decompositional technique to street-networks such that no network edge is longer than a set maximum distance, thereby increasing the resolution of observations (Figure~\ref{fig:decomposition}). A similar technique is exposed by the \code{cityseer-api} package: network edges can be `decomposed' so that no edge is longer than a specified distance. The effect is that topological artefacts at smaller distance thresholds can be forestalled while permitting increasingly precise assignment of data-points (e.g. land-uses or spatially embedded information) to adjacent street edges, thus encapsulating particularities at more finely spaced intervals along street-fronts. The overlapping nature of moving-window analysis means that the resolution of data sampling can be increased through decomposition without changing the spatial units of analysis. Decomposition is further discussed in the next section.

In \code{cityseer-api}, data aggregation is routed via the two street-network nodes on either side of the closest adjacent street edge, thus facilitating dynamic selection of the direction and distance of aggregation appropriate to the location of the currently windowed node. The assignment of data-points to adjacent streets is achieved with the use of a winding algorithm (Figure~\ref{fig:poi_assignment}), which first selects the closest adjacent node, then attempts to circle the street-network around the point of interest to identify the closest adjacent edge. If encountering a dead-end, the algorithm will backtrack and continue exploring. If exceeding the maximum search distance, it will then explore in the opposite winding direction to confirm whether any other closer edges exist.

\begin{figure}[htbp]
  \centering
  \includegraphics[width=\textwidth, keepaspectratio]{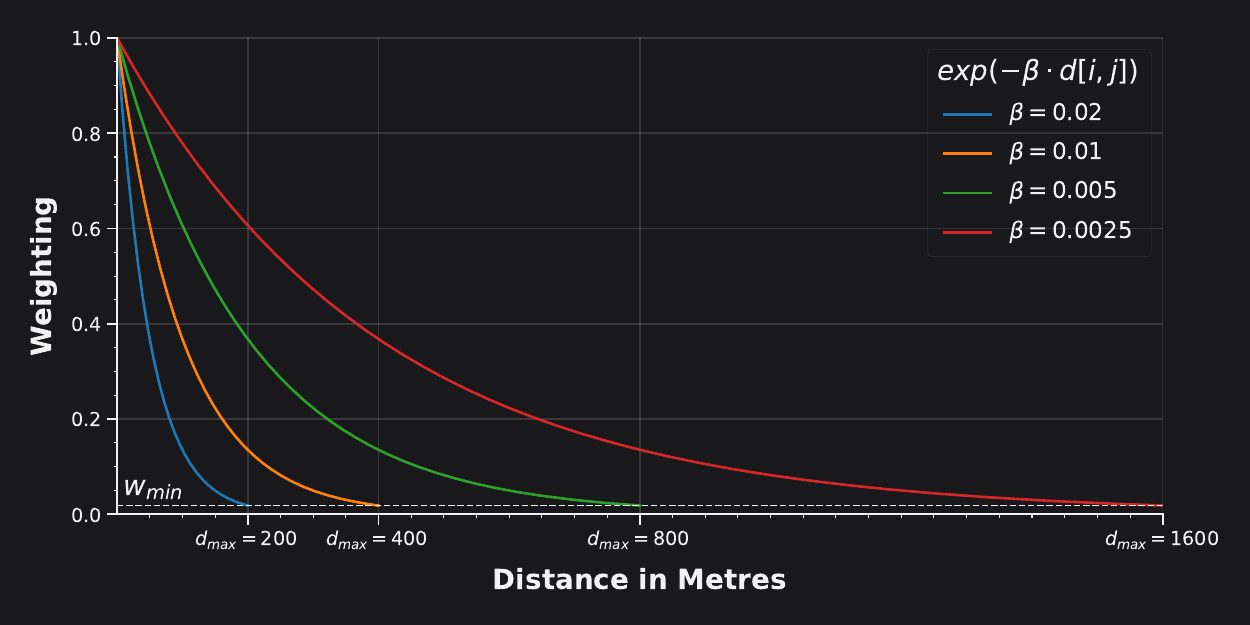}
  \caption[Spatial impedance curves for different $\beta$ parameters.]{Spatial impedance curves for different $\beta$ parameters. Nearer locations can be weighted more heavily than farther locations through use of the negative exponential decay (distance decay) function. The rate of the fall-off is controlled by the strength of $\beta$, where $\beta$ is by default selected as $4/d_{max}$. The mapping from $\beta$ to $d_{max}$ (or vice-versa) can be customised and is explained more comprehensively in the package documentation.}\label{fig:beta_decays}
\end{figure}

Since data-points remain located precisely in space, and distances are known from each windowed node to each data-point's location relative to the street-network and the direction of approach, the contribution of any such data-point can be weighted on an individual basis when calculating land-use accessibilities or other aggregative or statistical measures. Pedestrian walking-distance weighted aggregation methods applied over decomposed networks can, thus, yield a particularly localised emphasis while allowing for the use of equivalently larger distance thresholds that are less likely to incur topological artefacts otherwise encountered by small distance thresholds. Similar to spatial interaction models, \code{cityseer-api} models spatial impedances (distance decays) using the negative exponential decay function
\begin{equation}
  w = \exp(-\beta\cdot d)\, ,
\end{equation}
which reflects a decreasing willingness for pedestrians to walk correspondingly farther distances. Accordingly, the contribution of a data-point is weighted by weight $w$ as a function of the distance $d$. The rate at which this willingness to walk decreases is reflected in the strength of the specified $\beta$ parameter (Figure~\ref{fig:beta_decays}). By default, \code{cityseer-api} anchors the strength of $\beta$ relative to the selected maximum distance threshold of $d_{max}$ using
\begin{equation}
  \beta = 4 / d_{max}\, .
\end{equation}
This conversion can be manually specified where greater control is required over the relationship between $\beta$ and $d_{max}$; further information is provided in the documentation.

%% file: content/4_design_decisions.tex
\section{Design decisions}\label{design-decisions}

\subsection{Numba}

\code{cityseer-api} makes use of \code{Python} and \code{NumPy} \citep{Harris2020}, with computationally intensive algorithms optimised through use of \code{Numba} \code{JIT} compilation \citep{Lam2015}. \code{Python} is a widespread programming language offering access to an extensive ecosystem of high-quality network \citep[\code{networkX},][]{Hagberg2008}, geospatial \citep[\code{shapely},][]{Gillies2007}, OpenStreetMap conversion \citep[\code{OSMnx},][]{Boeing2017}, and data manipulation packages \citep[\code{pandas},][]{mckinney-proc-scipy-2010}. This facilitates general purpose workflows spanning from data munging and database I/O to interaction with an assortment of powerful data science and machine-learning packages such as \code{sklearn} \citep{Pedregosa2011} and \code{keras} \citep{chollet2015keras}.

The ease of use and flexibility of \code{Python} also entails a drawback: it offers slower performance when compared to lower-level languages such as \code{C}. For this reason, performance-critical \code{Python} packages are typically wrappers of code developed in more performant languages, with a prevalent example being the \code{NumPy} stack underpinning array-dependent operations central to a wide variety of \code{Python}'s scientific computing packages. Network-based methods, which depend on loop-intensive low-level algorithms such as Dijkstra's shortest path, remain a challenge, and it is for these purposes that the \code{Numba} package proves useful. \code{Numba} translates \code{Python} code into machine code using `Just In Time' (\code{JIT}) compilation, offering performance similar to that of compiled lower-level languages such as \code{C}. Use of \code{Numba} thus infers convenient access to the \code{Python} ecosystem while facilitating experimentation with computationally complex algorithms.

\subsection{Package composition}

The \code{cityseer-api} package consists of three sub-packages: \code{algos}, consisting of the \code{Numba} optimised functions; \code{metrics}, consisting of higher-level \code{Python} code accessed by the end-user; and \code{tools}, a collection of utility modules for purposes such as network preparation using the \href{https://cityseer.benchmarkurbanism.com/tools/graphs}{cityseer.tools.graphs} module and the generation and visualisation of mock data used by the unit tests.

The data structures and algorithms utilised by the \code{algos} sub-package's modules can be accessed directly; however, it is simpler to interact through the higher-level wrappers in the \code{metrics} sub-package. This contains the \code{networks} module for building street-networks and calculating street-network centralities as documented at \href{https://cityseer.benchmarkurbanism.com/metrics/networks/}{cityseer.benchmarkurbanism.com/metrics/networks/}. \code{cityseer-api} also broaches themes on land-use accessibilities, the mix of land-uses, and statistical aggregations, with these measures computed using the \code{layers} module as documented at \href{https://cityseer.benchmarkurbanism.com/metrics/layers/}{cityseer.benchmarkurbanism.com/metrics/layers/}. Whereas crude forms of these measures could be calculated using crow-flies distance aggregation methods, these become problematic when working at smaller pedestrian distance thresholds because the network structure can substantially affect distances to surrounding locations. The \code{layers} module is consequently underpinned by the same network structures and moving-window workflows utilised by the \code{networks} module. When functions contained in the \code{layers} module are invoked, data-points will be assigned to the specified street-network by invoking an algorithm that assigns each data-point to the closest adjacent street edge (see Figure~\ref{fig:poi_assignment}). Metrics computed by the \code{networks} and \code{layers} modules are computed relative to the same network structure, with calculations saved to a \code{GeoPanadas} \code{DataFrame} where they can be used for downstream statistical or machine-learning analysis.

An important advantage to network-based distance methods and the bidirectional assignment of data-points to network nodes is that comparatively accurate distances are known from any selected network node to any accessible data-point, thus allowing distance-weighted methods to be applied. These techniques are explored in more detail in the accompanying papers on network centrality methods \citep{Simons2021c} and mixed-use methods \citep{Simons2021d}.

\subsection{Decomposition}

A conundrum presents when calculating metrics on either a primal or dual network: architects, urban designers, and urbanists are interested in fine-scaled properties of the urban environment and how these properties can vary along street lengths. For example, characteristics at either street corner of a street segment may be notably different from that of the midpoint. One strategy may involve the interpolation of metrics to intervening locations, but this can be problematic for similar reasons; for example, if either end of a street segment has higher mixed-uses than the midpoint, interpolation will still give misleading results. The \code{cityseer-api} package therefor incorporates the optional use of network decomposition \citep{Yamada2010}. Each segment (edge) can be decomposed to a set maximum length (Figure~\ref{fig:decomposition}) so that longer street segments are broken down into smaller sections. This strategy confers some advantages when working at small distance thresholds: measurement can be performed at a higher-resolution of analysis and becomes more contextual; data-points can be assigned to the network more precisely; and longer street segments are no longer problematic if intersecting a distance threshold. Whereas the decomposed version entails additional computational demands, the benefit is a greater number of sampled points at a finer resolution.

From the perspective of classic forms of network analysis --- such as social networks, economic networks, web URL links, or citation networks --- the idea of decomposition may seem nonsensical. However, this works for urban analysis because the nodes and edges are not being used as fundamentally discrete units of analysis in the same sense of individual persons, businesses, URLs, or publications. These are, instead, used in a murkier sense as proxies mapping to the adjacently accessible street-network. As such, these forms of urban analysis are not, per se, about `intersections' as discrete points in space, but are instead about the availability of street frontages to pedestrians in a more continuous sense, and the resultant potential for social and economic activities as a function of the configuration of the street network. This intuition is borne out in a conceptual sense by Space Syntax's \citep{Hillier1984} use of the dual instead of the primal network representation, and in a practical sense by the use of street-length weighted \citep{Turner2007} or building weighted analysis \citep{Sevtsuk2012}. In short, longer streets or more granular street-front typologies represent larger generators of activity.

When applying network decomposition within the context of urban analysis, measures such as land-use accessibilities and mixed-uses increase in resolution and become more accurate but are not otherwise substantially affected because these are simple distance-weighted aggregations from a selected point on the street-network. The benefit of decomposition is that distance calculations and land-use assignments become increasingly precise, and the number of intervening nodes otherwise does not affect the calculations. Topological network centrality measures, such as network cycles, are likewise not greatly affected, though are now sampled at more frequent intervals on the street-network. On the other hand, aggregative node-based network centrality measures, such as closeness centrality and betweenness centrality, do behave differently on decomposed networks but not necessarily in a problematic sense:
\begin{itemize}
      \item Additional nodes result in additional summations, with the implication that the measures are not comparable across different levels of decomposition;
      \item Decomposition behaves as an implicit form of length-normalisation akin to weighting by street lengths because longer segments will yield a greater number of decomposed nodes, and therefor a greater number of summations. This effect is beneficial because nodes will be spread more evenly across the network, thereby tempering distortions introduced by varying concentrations of nodes on messier network representations;
      \item Decomposition introduces $degree=2$ nodes, and this can result in changes in the output distributions for aggregative centrality measures when calculated for small distance thresholds less than $200m$. See \citet{Simons2021c} for further discussion.
\end{itemize}

An alternative to node-based centrality measures is segmentised (continuous) forms of centrality adapted from their node-based equivalents. These are explored in \citet{Simons2021c}.
Segmentised measures explicitly acknowledge street-networks as a continuous rather than discretised form of analysis and remain stable when the network is decomposed; however, as with node-based measures, the distributions of the measures are affected by the introduction of $degree=2$ nodes for small distance thresholds.

%% file: content/5_prototypical_workflow.tex
\section{Example workflows}\label{prototypical-workflow}

Example workflows are provided and maintained on the \href{https://cityseer.benchmarkurbanism.com/examples}{examples} page of the documentation, with links to \code{Jupyter} notebooks.

Current examples include:
\begin{itemize}
  \item A getting started guide.
  \item An example workflow for cleaning and preparing OpenStreetMap data for analysis.
  \item An example for how OpenStreetMap data can be imported and converted from \code{OSMnx}.
  \item A demonstration showing how to compute network centralities for London.
  \item A demonstration showing how to compute pub accessibility for London.
\end{itemize}

\subsection{Graph Cleaning and Preparation}

\begin{figure}[htbp]
  \centering
  \begin{subfigure}[b]{0.45\textwidth}
    \centering
    \includegraphics[width=\textwidth]{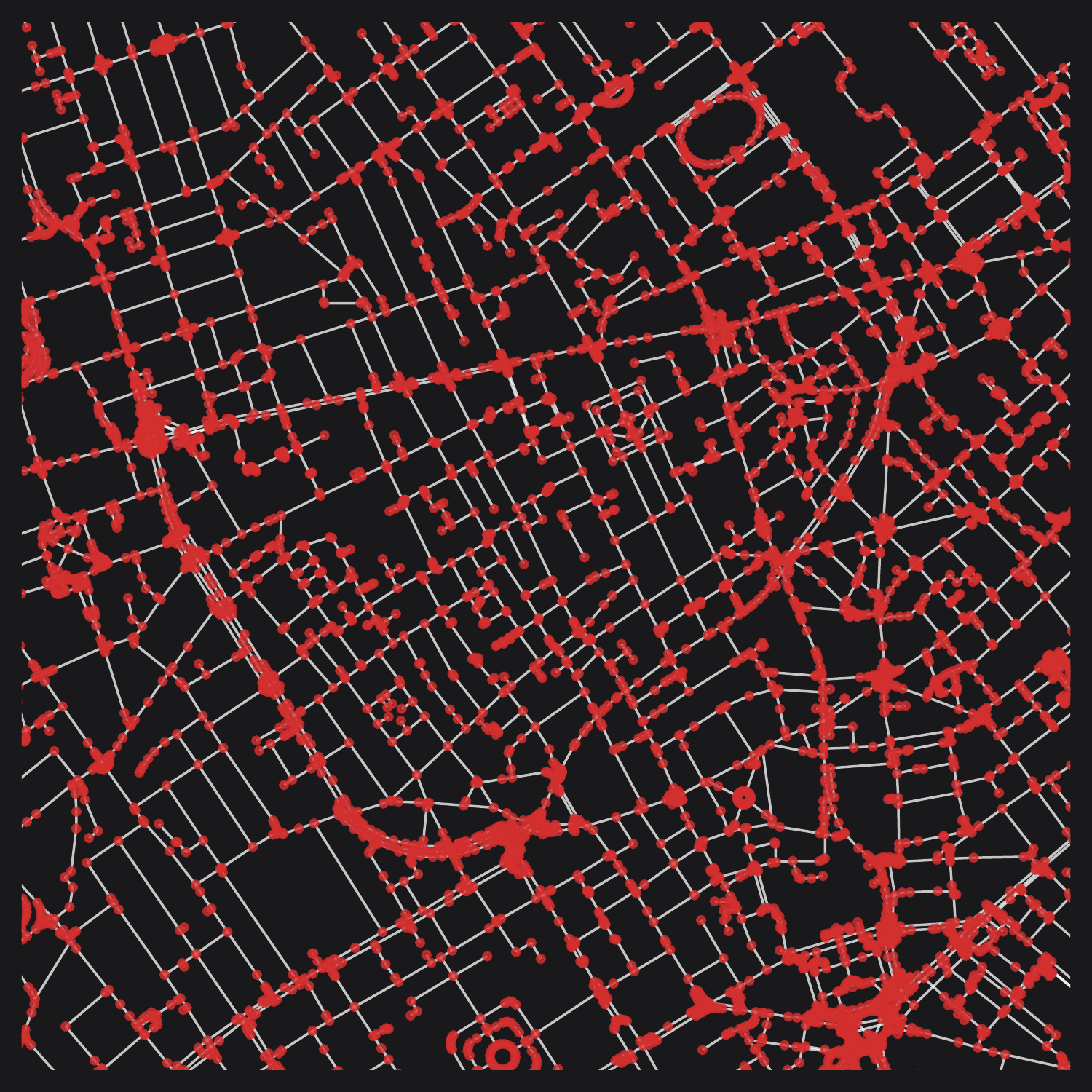}
    \caption{Raw network data as downloaded from OpenStreetMap prior to cleaning.}\label{cleaning-1}
  \end{subfigure}
  \begin{subfigure}[b]{0.45\textwidth}
    \centering
    \includegraphics[width=\textwidth]{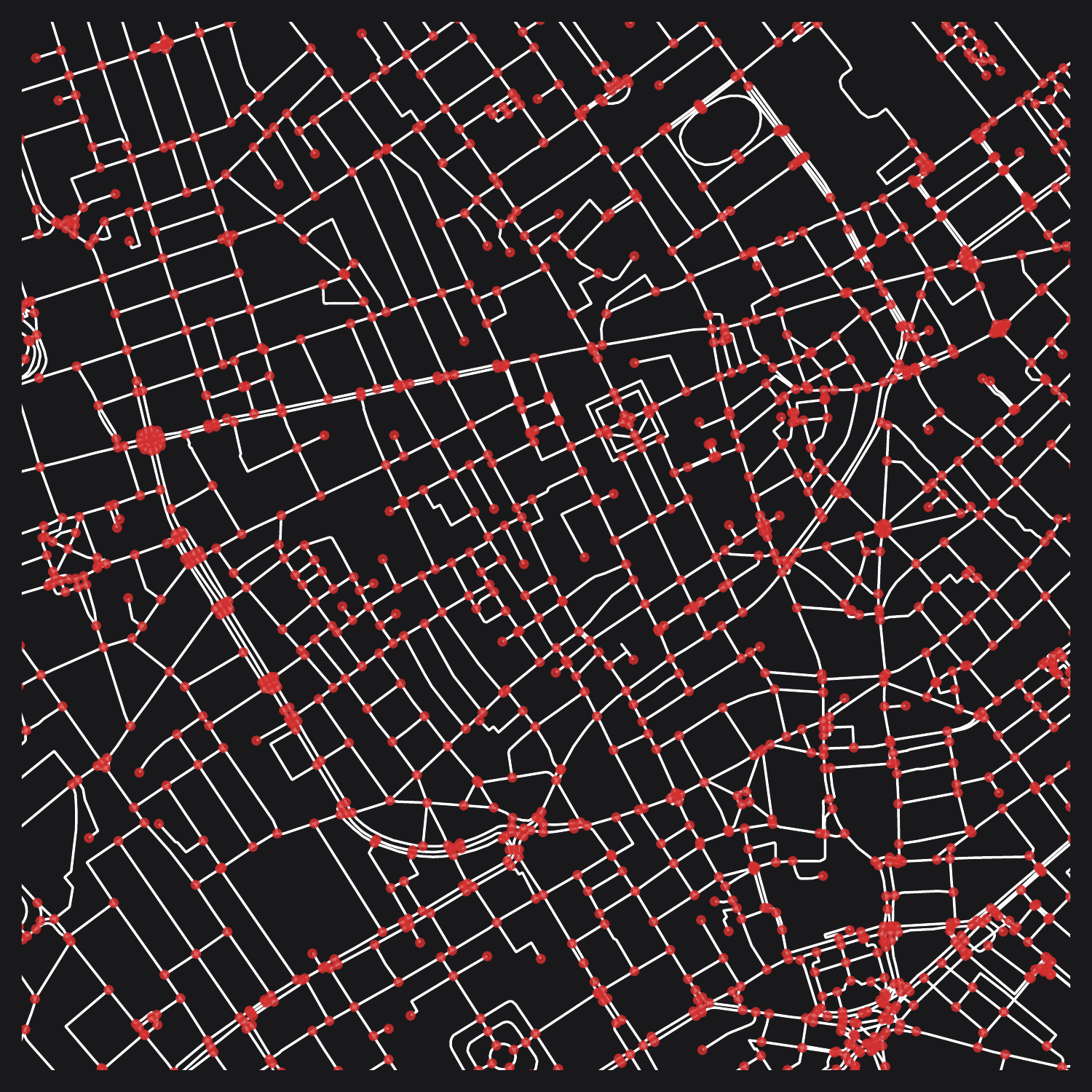}
    \caption{Graph data after removal of redundant `filler' and `dangling' nodes.}\label{cleaning-2}
  \end{subfigure}
  \begin{subfigure}[b]{0.45\textwidth}
    \centering
    \includegraphics[width=\textwidth]{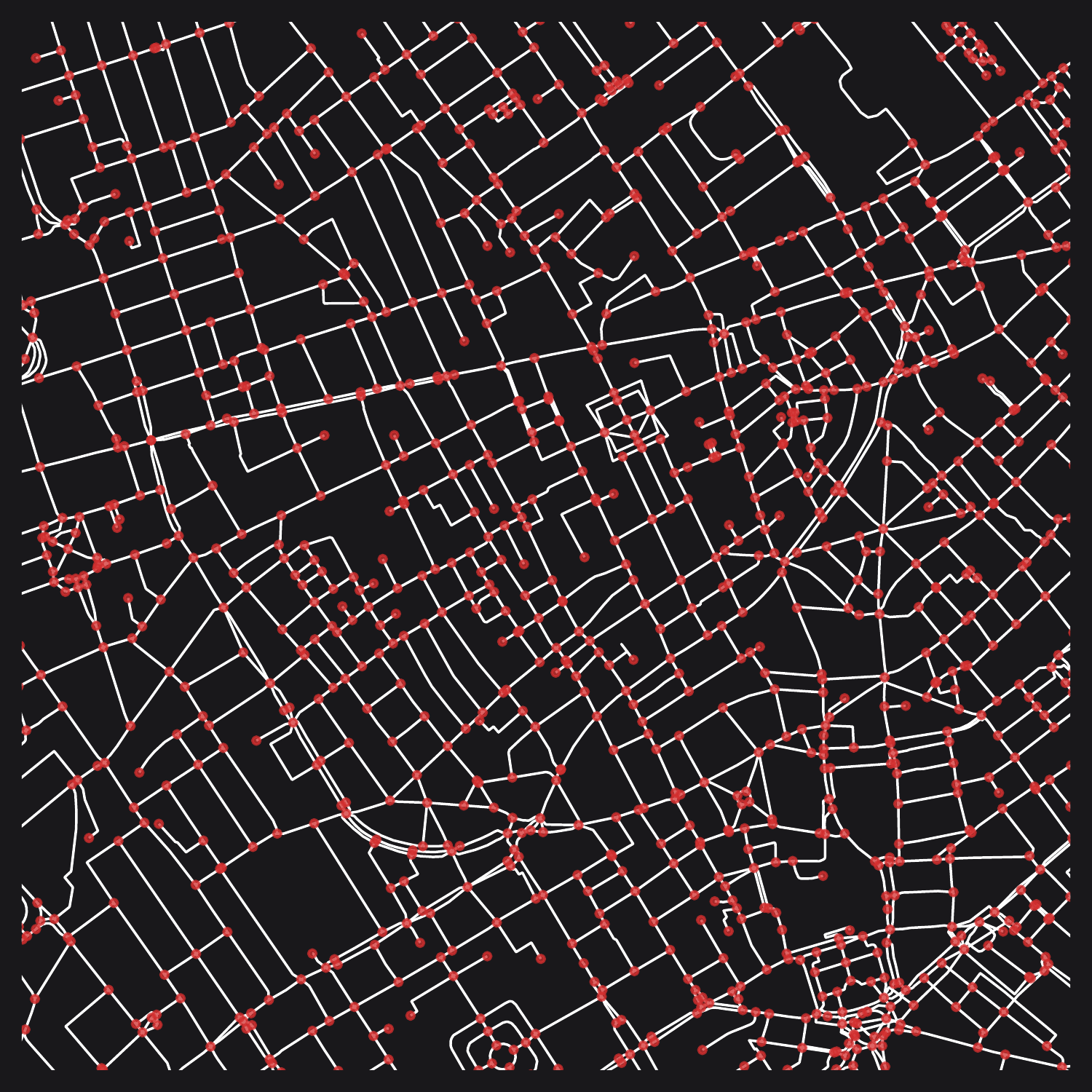}
    \caption{After an initial pass of network consolidation showing simplified intersections.}\label{cleaning-3}
  \end{subfigure}
  \begin{subfigure}[b]{0.45\textwidth}
    \centering
    \includegraphics[width=\textwidth]{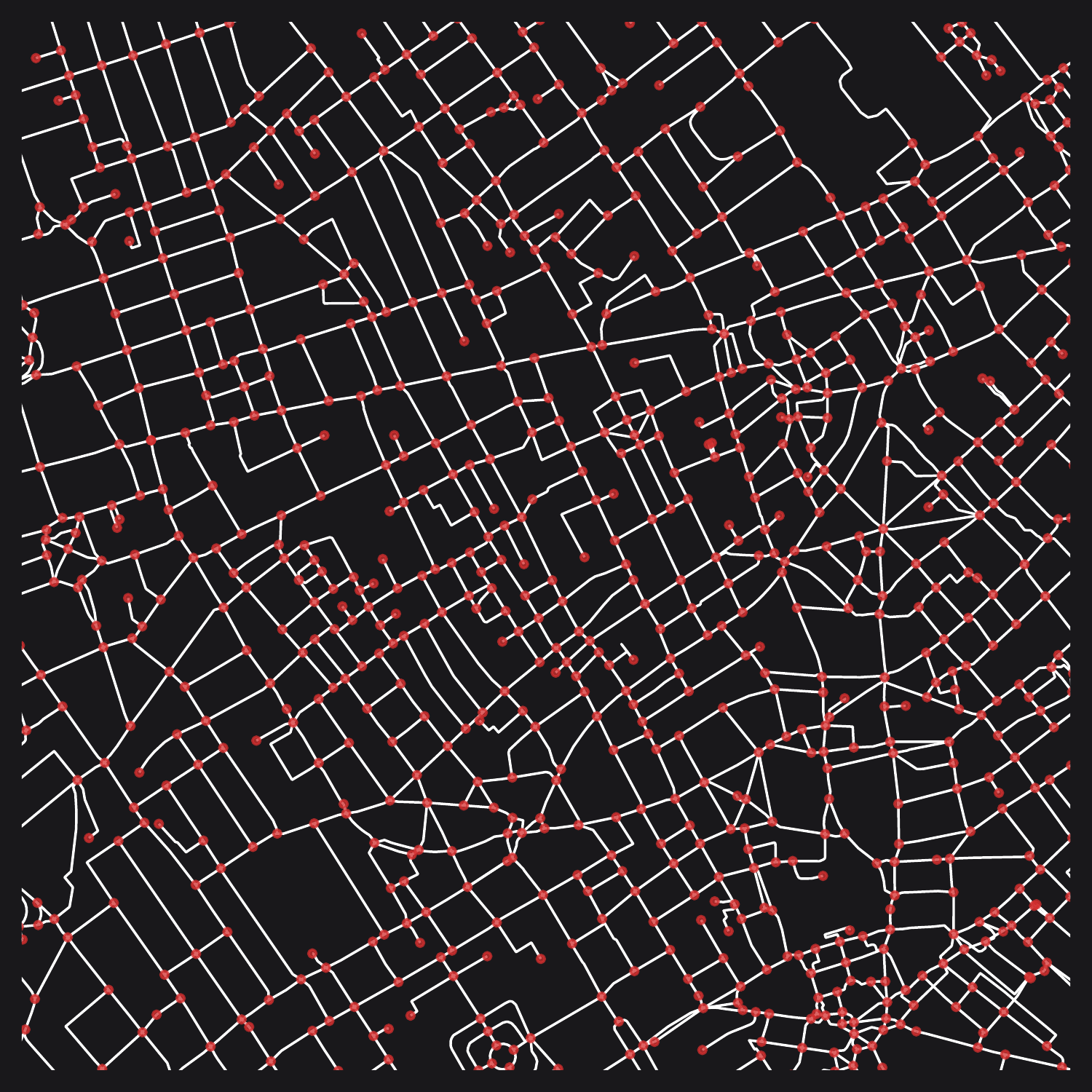}
    \caption{After a second pass of consolidation with parallel roadways removed.}\label{cleaning-4}
  \end{subfigure}
  \caption{Examples from typical network cleaning and preparation workflows. See the \href{https://cityseer.benchmarkurbanism.com/examples}{documentation examples} for links to \code{Jupyter} notebooks.}\label{network-cleaning}
\end{figure}

The \href{https://cityseer.benchmarkurbanism.com/tools/graphs/}{\code{tools.graphs}} module contains a collection of convenience functions for the preparation and conversion of \code{networkX} MultiGraphs, i.e.\ undirected networks allowing for multiple edges in cases where divergent street segments connect the same nodes. These functions are designed to work with raw \code{shapely} \href{https://shapely.readthedocs.io/en/latest/manual.html#linestrings}{\code{Linestring}} geometries assigned to edges as \code{geom} attributes. The benefit of this approach is that the geometry of the network remains decoupled from the topology: the topology is consequently free from distortions in node intensities which would otherwise confound centrality measures.

When creating a street-networks for analysis, two common scenarios might include:

\begin{itemize}
  \item Source datasets that keep the topology of the network separate from a street's geometry. This is the ideal case and the network can be constructed directly from the topology while assigning the roadway geometries to the respective edges spanning the nodes. \emph{Ordnance Survey} \href{https://www.ordnancesurvey.co.uk/business-and-government/products/os-open-roads.html}{\emph{Open Roads}} is an example of this type of dataset. Assigning the geometries to an edge involves firstly casting the geometry to a \code{shapely} \href{https://shapely.readthedocs.io/en/latest/manual.html#linestrings}{\code{Linestring}}, then assigning this geometry to the respective edge as a `\code{geom}' attribute.\ i.e.\ \code{G[start_node][end_node][edge_idx]['geom'] = linestring_geom}.
  \item Data sources that represent roadway geometries by adding additional nodes to the topological network. This is not desirable because this technique introduces topological distortions. In these cases, the \href{https://cityseer.benchmarkurbanism.com/guide#graph-cleaning}{\code{Graph Cleaning}} guide should be followed: the \href{https://cityseer.benchmarkurbanism.com/tools/graphs#nx-simple-geoms}{\code{graphs.nx_simple_geoms}} function can be used to generate street geometries and then several functions can be applied to further clean and prepare the network for analysis, including \href{https://cityseer.benchmarkurbanism.com/tools/graphs#nx-wgs-to-utm}{\code{nx_wgs_to_utm}} for WGS to UTM coordinate conversions; \href{https://cityseer.benchmarkurbanism.com/tools/graphs#nx-remove-dangling-nodes}{\code{nx_remove_dangling_nodes}} to remove roadway stubs and disconnected portions of the network, \href{https://cityseer.benchmarkurbanism.com/tools/graphs#nx-remove-filler-nodes}{\code{nx_remove_filler_nodes}} to strip out unnecessary filler nodes, and \href{https://cityseer.benchmarkurbanism.com/tools/graphs#nx-consolidate-nodes}{\code{nx_consolidate_nodes}} to consolidate nodes.
\end{itemize}

Related examples are provided in the documentation \href{https://cityseer.benchmarkurbanism.com/examples}{examples}, with example images shown in Figure~\ref{network-cleaning}.

\subsection{Computationally Efficient Analysis}

\begin{figure}[htbp]
  \centering
  \begin{subfigure}[b]{0.425\textwidth}
    \centering
    \includegraphics[width=\textwidth]{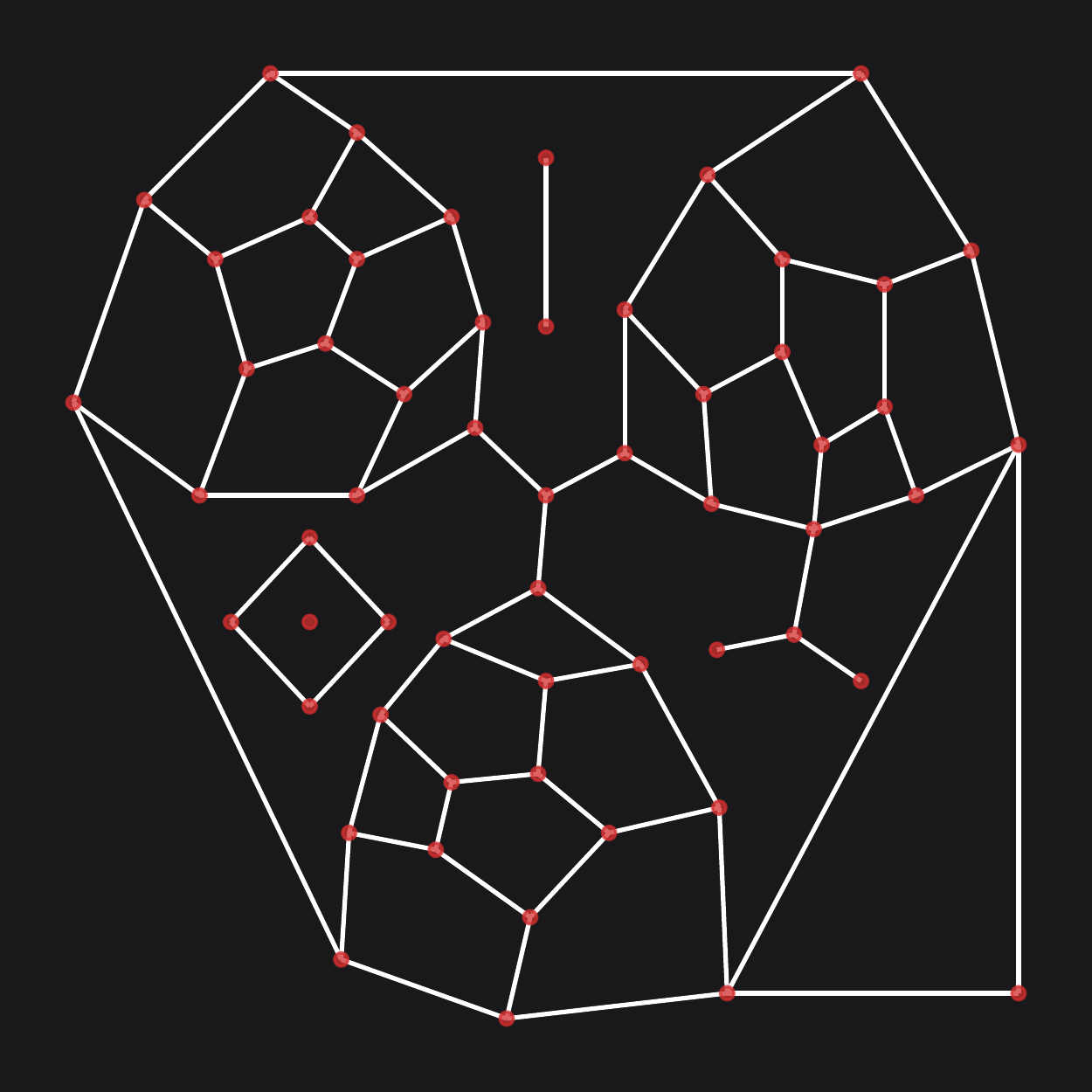}
    \caption{A mock network for demonstration.}\label{network-simple}
  \end{subfigure}
  \begin{subfigure}[b]{0.425\textwidth}
    \centering
    \includegraphics[width=\textwidth]{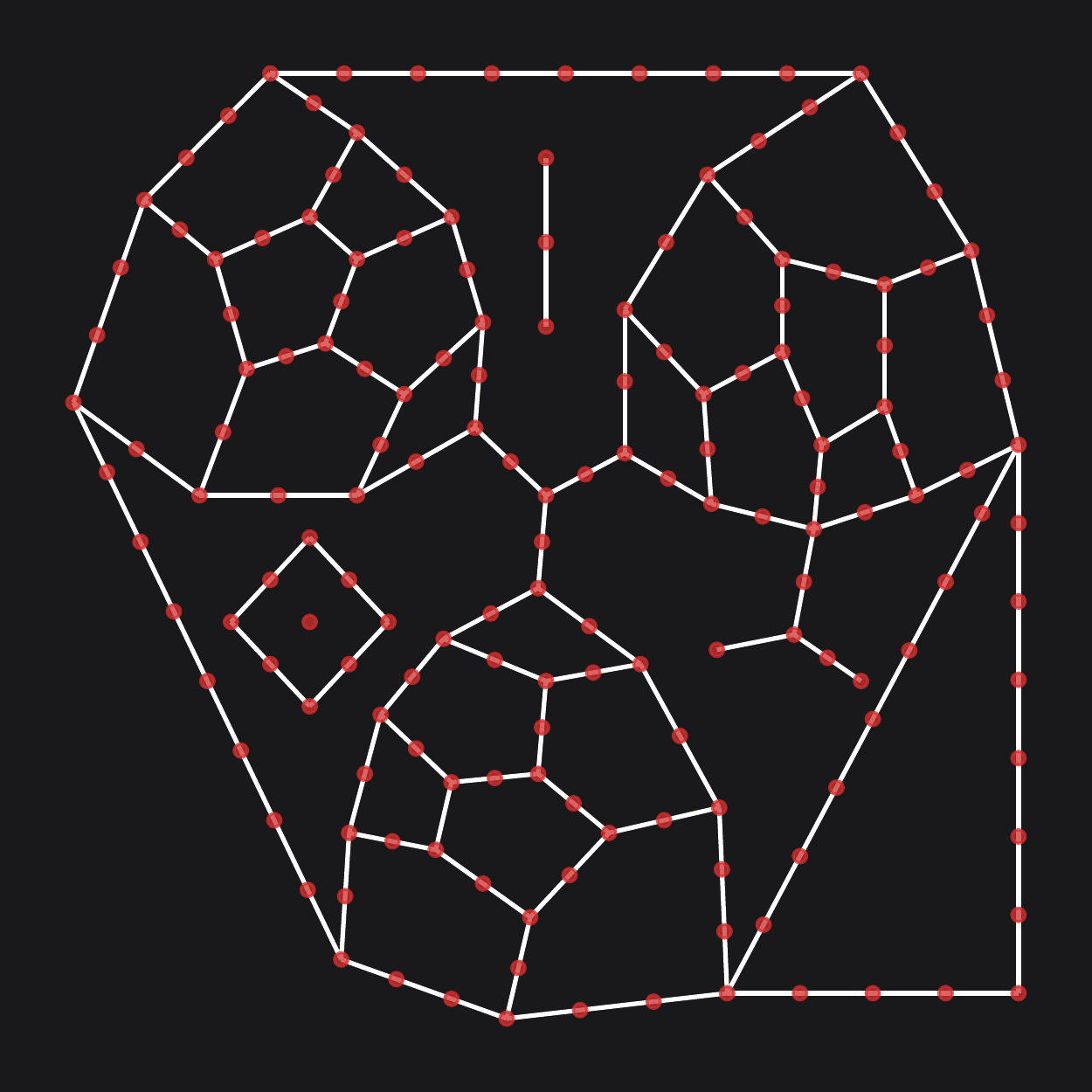}
    \caption{The network after decomposition.}\label{network-decomposed}
  \end{subfigure}
  \begin{subfigure}[b]{0.425\textwidth}
    \centering
    \includegraphics[width=\textwidth]{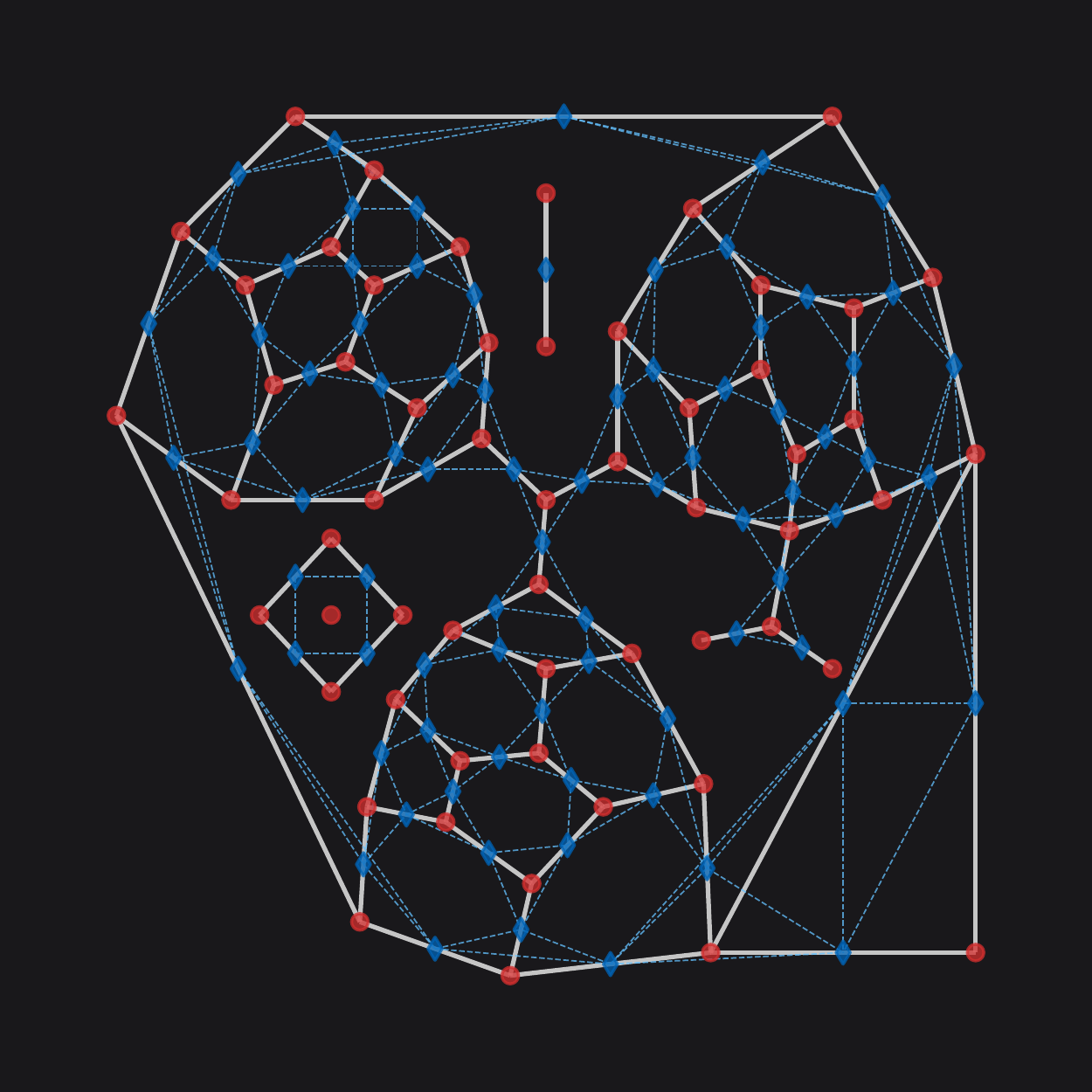}
    \caption{The network after primal to dual conversion. Street geometries are accordingly sliced and welded to form the new geometries.}\label{network-dual}
  \end{subfigure}
  \begin{subfigure}[b]{0.425\textwidth}
    \centering
    \includegraphics[width=\textwidth]{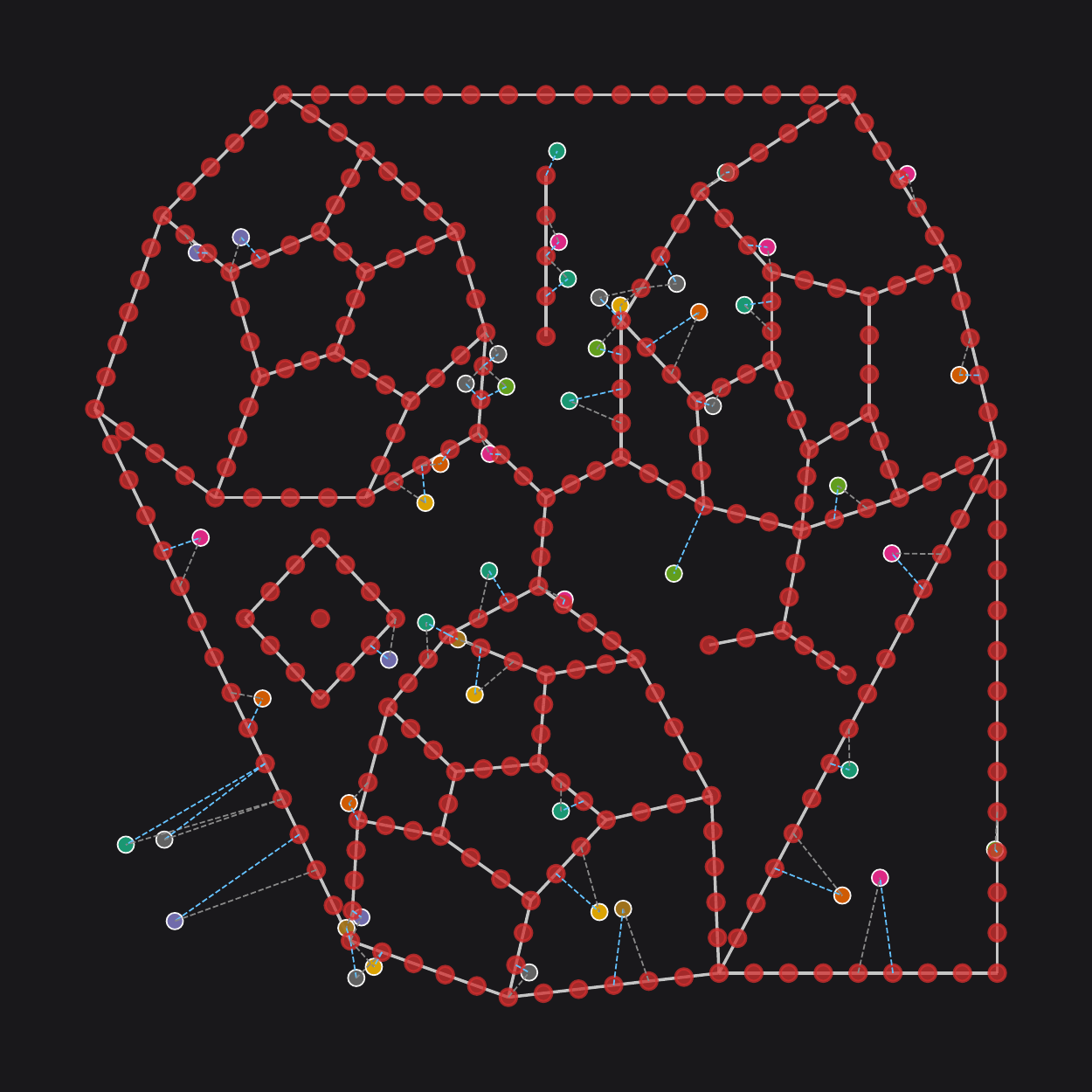}
    \caption{Bi-directional assignment of data-points to adjacent edges. Aggregations use distances calculated relative to the direction of approach.}\label{network-assignment}
  \end{subfigure}
  \begin{subfigure}[b]{0.425\textwidth}
    \centering
    \includegraphics[width=\textwidth]{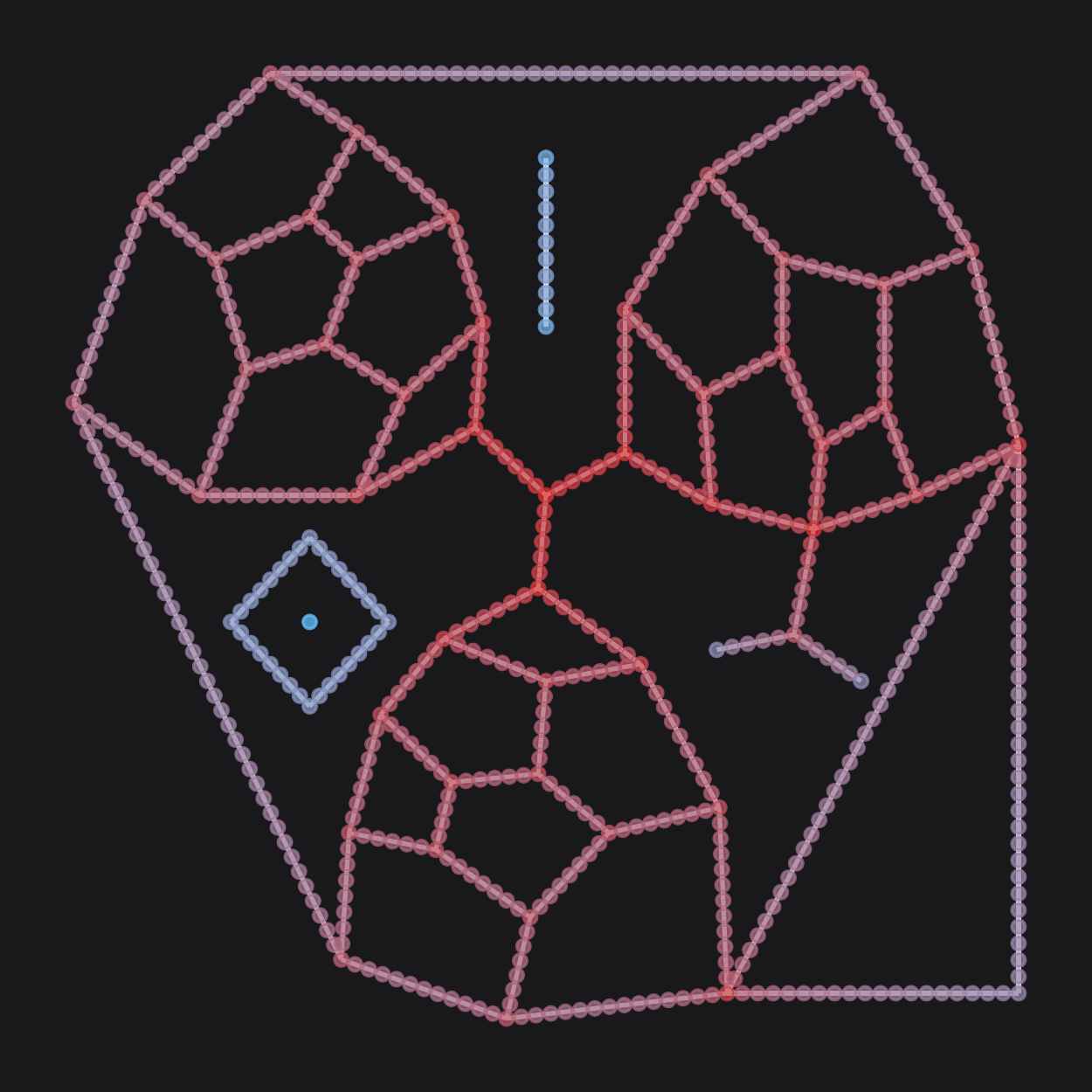}
    \caption{$800m$ segmentised harmonic centrality.}\label{segmentised-harmonic-centrality}
  \end{subfigure}
  \begin{subfigure}[b]{0.425\textwidth}
    \centering
    \includegraphics[width=\textwidth]{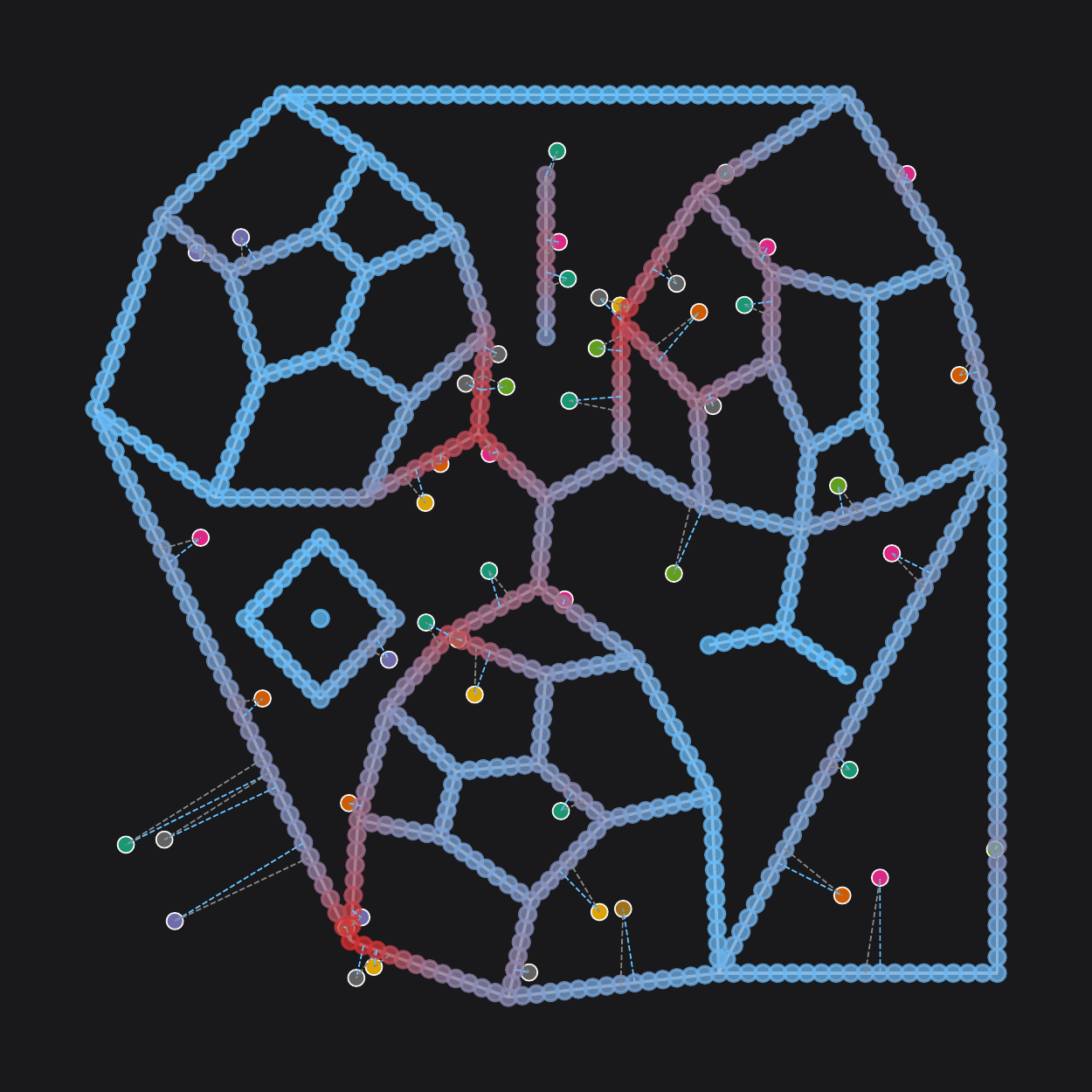}
    \caption{$400m$ distance-weighted mixed-uses.}\label{distance-weighted-mixed-uses}
  \end{subfigure}
  \caption{See the \href{https://cityseer.benchmarkurbanism.com/examples}{documentation examples} for links to \code{Jupyter} notebooks.}\label{landuse-and-stats}
\end{figure}

After network preparation and cleaning is completed, the \code{networkX} network can be transformed into the data structures used by the \code{networks} and \code{layers} modules for efficiently computing centralities, land-use measures, and statistical aggregations. This is done by calling the \href{https://cityseer.benchmarkurbanism.com/tools/graphs#network-structure-from-nx}{\code{network_structure_from_nx}} function to convert a \code{networkX} network into a \code{GeoPandas} \href{https://geopandas.org/en/stable/docs/reference/api/geopandas.GeoDataFrame.html}{\code{GeoDataFrame}} representing the data state for each node, and a \href{https://cityseer.benchmarkurbanism.com/structures#networkstructure}{\code{structures.NetworkStructure}} containing detailed information of the network for use by underlying algorithms.

The \href{https://cityseer.benchmarkurbanism.com/metrics/networks#node-centrality}{\code{networks.node_centrality}} and \href{https://cityseer.benchmarkurbanism.com/metrics/networks#segment-centrality}{\code{networks.segment_centrality}} methods wrap underlying \code{Numba} optimised functions for computing a range of available centrality methods. Specified measures and distance thresholds are computed simultaneously to reduce the time required for multi-variable and multi-scalar workflows. The results of the computations are written to the nodes \code{GeoDataFrame} for downstream analysis.

Land-use and statistical measures require a \code{GeoPandas} \code{GeoDataFrame} representing data-points. \code{cityseer-api} automatically routes the location of each data-point through the two closest network nodes, one in either direction, as determined from the closest adjacent street edge. This permits \code{cityseer-api} to use dynamic spatial aggregation methods that more accurately describe distances from the perspective of pedestrians travelling over the network, and relative to the direction of approach.

The \href{https://cityseer.benchmarkurbanism.com/metrics/layers#compute-landuses}{\code{layers.compute_landuses}} function is used for the calculation of mixed-use and land-use accessibility measures. In this case, \code{GeoDataFrame} columns are used to represent categorical land-use information (e.g. `pub', `shop', `school'). As with the centrality methods, land-use measures are computed simultaneously for all selected forms of analysis; however, stand-alone methods are also available, including \href{https://cityseer.benchmarkurbanism.com/metrics/layers#hill-diversity}{\code{layers.hill_diversity}}, \href{https://cityseer.benchmarkurbanism.com/metrics/layers#hill-branch-wt-diversity}{\code{layers.hill_branch_wt_diversity}}, and \href{https://cityseer.benchmarkurbanism.com/metrics/layers#compute-accessibilities}{\code{layers.compute_accessibilities}}.

The \href{https://cityseer.benchmarkurbanism.com/metrics/layers#compute-stats}{\code{layers.compute_stats}} function is used for statistical aggregations. In this case, \code{GeoDataFrame} columns are used to represent numerical information.

Land-use metrics and statistical aggregations are computed over the street-network relative to the network, with results written to each node. The mixed-use, land-use accessibility, and statistical aggregations can therefor be compared directly to centrality computations performed from the same locations, which can then be fed to downstream statistical or machine-learning analysis. Data derived from the \code{cityseer.metrics} package can be converted back into a \code{NetworkX} network using the \href{https://cityseer.benchmarkurbanism.com/tools/graphs#nx-from-network-structure}{\code{nx_from_network_structure}} function, which can also overlay computed metrics onto the original network if provided as a \code{nx_multigraph} parameter to the function.

Related examples are provided in the documentation \href{https://cityseer.benchmarkurbanism.com/examples}{examples}, with example images shown in Figure~\ref{landuse-and-stats}.

%% file: content/6_summary.tex
\section{Summary}

\code{cityseer-api} contributes a synthesis of computational techniques to support granular forms of network-based spatial analysis from the perspective of pedestrians:
\begin{itemize}
      \item High-resolution workflows using localised moving-window analysis with strict network-based walking distance thresholds; spatially precise assignment of land-uses or other data-points to adjacent street-fronts for improved contextual sensitivity; dynamic aggregation workflows which aggregate data-points and compute distances on-the-fly from any selected point on the network to any accessible land-use or data-point within a selected distance threshold; facilitation of workflows eschewing intervening steps of aggregation and associated issues such as ecological correlations; and the optional use of network decomposition to increase the resolution of the analysis.
      \item Computation of network centralities using either shortest or simplest path heuristics on either primal or dual networks, including tailored methods such as harmonic closeness centrality, which behaves more suitably than traditional globalised variants of closeness, and segmentised versions of centrality, which convert centrality methods from a discretised to an explicitly continuous form, see \citet{Simons2021c}.
      \item Land-use accessibilities and mixed-use calculations incorporate dynamic and directional aggregation workflows with the optional use of spatial-impedance-weighted forms. These can likewise be applied with either shortest or simplest path heuristics and on either primal or dual networks, see \citet{Simons2021d}.
      \item Network centralities dovetailed with land-use accessibilities, mixed-uses, and general statistical aggregations from the same points of analysis to generate multi-scalar and multi-variable datasets facilitating downstream data science and machine-learning workflows, see \citet{Simons2021} and \citet{Simons2021a}.
      \item The inclusion of network cleaning methods reducing topological distortions for high quality network analysis and aggregation workflows while accommodating workflows bridging the wider \code{Numpy} and \code{GeoPandas} ecosystem of scientific and geospatial packages.
      \item \code{Numba} \code{JIT} compilation of underlying loop-intensive algorithms allows for these methods to be applied to large and, optionally, decomposed networks, which have greater computational demands.
\end{itemize}

%% file: shared/acknowledge_phd.tex
\subsection{PhD}

This paper derives from the author's PhD research at the \emph{Centre for Advanced Spatial Analysis}, \emph{University College London}. The author wishes to acknowledge their PhD supervisors, Prof.~Elsa Arcaute and Prof.~Michael Batty, for their gracious support and feedback throughout the development of this work. The author takes sole responsibility for any oversights or shortcomings contained within this paper.